\RequirePackage{amsmath}

\documentclass[runningheads]{llncs}

\usepackage{amscd, epsfig}
\usepackage{graphicx}
\usepackage{multirow, array}
\usepackage{color, colortbl}
\usepackage[misc]{ifsym}

\newcolumntype{x}[1]{>{\centering\arraybackslash\hspace{0pt}}p{#1}}
\newcommand{\comment}[1]{}

\begin{document}

\title {A Soft STAPLE Algorithm Combined with Anatomical Knowledge}

\author{
Eytan Kats\inst{1} \and 
Jacob Goldberger\inst{2} \and
Hayit Greenspan\inst{1}
}


\authorrunning{E. Kats et al.}

\institute{Tel-Aviv University, Tel-Aviv, Israel \\
\email{eytankats@mail.tau.ac.il} \\
\and Bar-Ilan University, Ramat-Gan, Israel}

\maketitle

\begin{abstract}

Supervised machine learning algorithms, especially in the medical domain, are affected by considerable ambiguity in expert markings. In this study we address the case where the experts' opinion is obtained as a distribution over the possible values. We propose a soft version of the STAPLE algorithm for experts' markings fusion that can handle soft values. The algorithm was applied to obtain consensus from soft Multiple Sclerosis (MS) segmentation masks. Soft MS segmentations are constructed from manual binary delineations by including lesion surrounding voxels in the segmentation mask with a reduced confidence weight. We suggest that these voxels contain additional anatomical information about the lesion structure. The fused masks are utilized as ground truth mask to train a Fully Convolutional Neural Network (FCNN). The proposed method was evaluated on the MICCAI 2016 challenge dataset, and yields improved precision-recall tradeoff and a higher average Dice similarity coefficient.
  
  
\end{abstract}

\section{Introduction}
\label{sec:introduction}

Manual analysis of medical data is liable to  inter-expert performance variability due to differences of interpretation and level of expertise. Supervised machine learning algorithms for detection and segmentation in the medical domain are affected by considerable ambiguity in the expert annotations. The automatic segmentation task of MS lesions is especially challenging since lesion contours are not well defined on MRI images which leads  to considerable  ambiguity in the expert markings along the lesion contours. Accurate segmentation of MS lesions is essential for reliable disease onset detection, when tracking its progression and in evaluating treating efficiency. This makes it crucial to train the model on the most likely labels that are determined by fusing the annotations of different experts. A principled way to address the annotations fusion problem is to build generative probabilistic models of the expert decision processes, and assign labels using standard inference tools. The expert reliability is viewed as an unknown parameter. Several  works applied the EM algorithm to this task by incorporating either simple or more complicated generative models  (e.g. \cite{accuracy_of_labeling_sources_donmez,ground_truth_from_subjective_labelling_smyth,integration_of_labels_whitehill,wisdom_of_crowds_welinder,eliminating_spammers_raykar}). The best known approach in medical imaging is  STAPLE (Simultaneous truth and performance level estimation) \cite{staple_warfield,staple_tutorial_alireza}.

In this study we address the problem of combining ground truth labeling from several human annotators who assign a soft (lesion, non-lesion) value to each image voxel. In the classical STAPLE setup, the experts  provide  deterministic binary decisions. Here we  assume  that each expert splits his vote among the possible voxel labels. The opinion of an expert is thus provided in the  form of a distribution over the possible values.

In this paper we propose a modified STAPLE algorithm for experts' soft annotations and apply it to create a fusion of soft masks constructed from the manual binary MS lesions delineations using anatomical knowledge according to the protocol described in \cite{soft_labeling_kats}. We used the dataset of the MICCAI 2016 MS lesions segmentation challenge (MSSEG dataset) \cite{msseg_commowick} that contains seven manual delineations for each MS patient case. The  soft STAPLE  created a  soft consensus mask that takes advantage of  the anatomical knowledge of the  lesions structure. We show that training a Fully Convolutional Neural Network (FCNN) with the proposed soft consensus mask enhances the  performance  compared to the FCNN trained with the mask created by the classic STAPLE algorithm.

\section{A Modified STAPLE Algorithm for Experts' Soft Annotations}
\label{sec:soft_staple}

We start by reviewing the STAPLE algorithm for the simpler and standard case where the expert human annotators  provide binary 0/1 labeling, and then extend it to soft labeling.
Assume $x_1,...,x_n$ are  random binary variables. In the segmentation of MS lesions, the value of $x_i$ indicates whether the voxel is in a lesion area or not.
The values of  $x_1,...,x_n$ are not directly observed.  Instead, there is a set of $m$ `experts' and the opinion of expert $i$ on the value $x_t$ is denoted by $y_{it}\in \{0,1\}$.  We assume that
 each expert $i$ is associated with sensitivity and specificity parameters.
The sensitivity parameter of the $i$-th  is defined as $\theta_{i1}=p(y_{it}=1|x_t=1)$ and in a similar way the specificity parameter is defined as $\theta_{i0}=p(y_{it}=0|x_t=0)$.
 Let $y_t=\{y_{1t},...,y_{mt}\}$ be the experts' opinions on the value of $x_t$. Assuming the annotations are independently provided by the $m$ experts, the probability of the annotations of the $t$-th voxel is:
 \begin{equation}
  p(y_{t}|x_t=a;\theta) =  \prod_{i=1}^m p(y_{it}|x_t=a;\theta_{ia}),    \hspace{1cm} a \in\{0,1\}
  \label{condprobs}
 \end{equation}
such that  $\theta$ is the parameter set.  Given the experts' annotations we can compute the posterior distribution of  $x_t$. Applying Bayes' rule, we obtain:
  \begin{equation}
\label{estep}   p( x_t\!=1 | y_t; \theta) =
\frac{ p_{prior} p(y_t|x_t=1;\theta)}{ (1-p_{prior})p(y_t|x_t=0;\theta) + p_{prior}p(y_t|x_t=1;\theta)}
 \end{equation}
 where $p_{prior}$ is the prior probability of a voxel to be a lesion.

The goal of the STAPLE algorithm is to find both the expert reliability parameters and the ground truth segmentation
using the given expert information set $y_1,...,y_n$ of the $n$ voxels. The log-likelihood function is:
\begin{equation}
\begin{aligned}
 L(\theta)  &= \sum_{t=1}^n \log p(y_t;\theta)  =  \sum_{t=1}^n \log (  \sum_{a=0,1}   p(y_{t},x_t=a;\theta)   ).
 \end{aligned}
\label{probmodel}
\end{equation}

The EM algorithm handles the parameter estimation task
by iterating between the E and M steps.
  The E-step is:
\begin{equation}
\label{estep1}  w_t(a) = p( x_t\!=\!a | y_t; \theta),
  \hspace{1cm} t=1,...,n, \hspace{0.5cm}a\in \{0,1\}
\end{equation}
where $\theta$ is the current value of the parameter set and $p( x_t\!=\!a | y_t; \theta)$ is defined in Eq. (\ref{estep}).
The M-step is composed of updating the sensitivity and specificity parameters:
\begin{equation}
\theta_{i1}  =  \frac{\sum_{t=1}^n  y_{it} w_t(1)}{  \sum_{t=1}^n w_t(1)}, \hspace{0.5cm}
\theta_{i0}  =  \frac{\sum_{t=1}^n  (1-y_{it}) w_t(0)}{  \sum_{t=1}^n w_t(0)},
\hspace{1cm} i=1,...,m.
\label{m_step1}
\end{equation}
After the algorithm  converges, we can extract a binary labeling from Eq. (\ref{estep}):
\begin{equation}
\hat{x}_t= \arg \max_{a\in\{0,1\}}  p( x_t\!=\!a | y_t; \theta), \hspace{1cm} t=1,...,n
\label{hard_em_hard_decision}
\end{equation}
that can be used as a ground truth for training a lesion segmentation network.

We next extend the problem of combining the opinions of several expert annotators to the case where the experts provide a soft opinions in the form of a distribution over the set of possible decisions (either 0 or 1).
The opinion of expert $i$ on the value of $x_t$ is thus provided in the  form of a distribution:
\begin{equation}
q_{it}(b) = p(y_{it}=b), \hspace{1.2cm}b\in \{0,1\}.
\end{equation}
Assuming the experts' opinions are independently generated, we use the following notation for the soft opinions on the binary value of the voxel $x_t$:
\begin{equation}
q_t(B) = \prod_{i=1}^m q_{it}(b_i),   \hspace{0.5cm} \mbox{s.t.} \hspace{0.5cm}  B=(b_1,...,b_m)\in \{0,1\}^m.
\label{qjdef}
\end{equation}
The modified cost function  we optimize here is:
\begin{equation*}
  L_{soft}(\theta) =  \sum_{t=1}^n E_{q_t} \log p( y_t)
  = \sum_{t=1}^n \sum_{B\in \{0,1\}^m}q_{t}(B)  \log ( \sum_{a\in \{0,1\}}   p(y_{t}=B,x_t=a;\theta)   ).
\end{equation*}

The optimal parameter can be found by a modification of the EM algorithm defined above.
The E-step is:
\begin{equation}
  w_t(a) = \sum_{B\in \{0,1\}^m}q_{t}(B)  p(x_t=a| y_t=B;\theta),
  \hspace{0.4cm} t=1,...,n, \hspace{0.4cm}a\in \{0,1\}
  \label{soft_E}
 \end{equation}
 such that $\theta$ is the current value of the parameter-set and
 \begin{equation*}
  p(x_t\!=\!a| y_t\!=\!B;\theta) =
  \frac{ p_{prior} p(y_t\!=\!B|x_t\!=\!1;\theta)}{ (1-p_{prior})p(y_t\!=\!B|x_t\!=\!0;\theta) + p_{prior}p(y_t\!=\!B|x_t\!=\!1;\theta)}
 \end{equation*}
where
 \begin{equation*}
 p(y_t=B|x_t=a;\theta) = \prod_{i=1}^m p(y_{it} = b_i | x_t=a;\theta).
 \end{equation*}

The M-step remains the same as above. The sensitivity and specificity  parameters are updated using (\ref{m_step1}).
It can be easily verified that this iterative algorithm monotonically increases $L_{soft}(\theta)$. 
Once we have found the model parameter-set $\theta$, we can use Eq. (\ref{soft_E}) to compute a soft ground truth labeling for each voxel  $x_t$.

Note that the complexity of computing the expressions $w_t(a)$  (\ref{soft_E}) in the E-step is exponential in the number of experts (see \cite{combining_soft_decisions_goldberger} for an approximation method). We next describe a simplified likelihood function with an easily computed  E-step. 
Consider the soft labeling as an observed  noisy version $z_{it}$ of the exact expert opinion $y_{it}$.
\begin{equation}
p(z_{it}|x_t=a;\theta) = \sum_{b=0,1} q_{it}(b)  p(y_{it}=b|x_t=a;\theta_{ia}),\hspace{0.5cm}a\in \{0,1\}
\end{equation}
i.e. $p(z_{it}|x_t=1;\theta) = q_{it}(1) \theta_{i1} + q_{it}(0) (1-\theta_{i1})$ and $p(z_{it}|x_t=0;\theta)=q_{it}(1) (1-\theta_{i0} + q_{it}(0) \theta_{i0}$.
Let $z_t=\{z_{1t},...,z_{mt}\}$ be the soft manual annotations regarding the value of $x_t$. The likelihood function here is:
\begin{equation}
 L_{simple}(\theta) =  \sum_{t=1}^n  \log p( z_t)
  = \sum_{t=1}^n \sum_{i=1}^m \log ( \sum_{a\in \{0,1\}}   p(z_{it},x_t=a;\theta)   ).
\end{equation}
The E-step here is easily computed:
 \begin{equation}
  w_t(1)= p(x_t\!=\!1| z_t;\theta) =
  \frac{ p_{prior} p(z_t|x_t\!=\!1;\theta)}{ (1-p_{prior})p(z_t|x_t\!=\!0;\theta) + p_{prior}p(z_t|x_t\!=\!1;\theta)}
 \end{equation}
where
 \begin{equation}
 p(z_t|x_t=a;\theta) = \prod_{i=1}^m p(z_{it}  | x_t=a;\theta).
 \end{equation}
 The M-step remains the same as above. 
 We dub the first label fusion algorithm the soft-STAPLE and the second  algorithm the simplified-soft-STAPLE. In the next section we show that the former algorithm yields better results.

\section{Soft Labeling by Anatomical Knowledge}
\label{sec:soft_labeling}

In this section we describe a situation where expert annotation is given in the form of soft labeling. 
In the framework of the MS lesions segmentation task, soft masks can be created following the protocol described in \cite{soft_labeling_kats}. This method uses the observation that most of the inter-rater variability in MS lesions manual delineations is found along the MS lesion contour voxels. The true delineations at these voxels can be extended by adding voxels with soft labels. In order to create the soft mask, the original binary mask is expanded by 3D morphological dilation. Using the clinical observation that lesions appear as hyper-intense regions in FLAIR images \cite{segmentation_with_spatial_consistency_mechrez}, those voxels with a FLAIR intensity value below a defined threshold are excluded from the dilated region. Selected voxels from the dilated region are assigned a soft label ${0<\gamma<1}$ which is interpreted as the  probability of the voxel being  part of the lesion. The label of the manually annotated voxels remains 1.
Lesion from the same tissue surrounding the marked pixels may also include some lesion level information. We  can thus create a soft labeling that can be used via the  Dice function to provides additional information about lesion structure during the training process beyond the ground truth mask obtained by the expert.

Training with imbalanced data is very problematic especially when the training evaluation measure is classification accuracy. A well-known alternative method for evaluating the performance of medical imaging systems is the Dice measure.
 The Dice loss function is defined as:
\begin{equation}
\label{dice_loss}
  Dice Loss 
  = -\frac{TP}{TP + 0.5FP + 0.5FN} =
  -\frac{\sum_{i} (\textbf{T}_{i} \cdot \textbf{P}_{i})}{0.5\sum_{i} \textbf{P}_{i} + 0.5\sum_{i} \textbf{T}_{i}}
\end{equation}
where TP is the number of True Positive voxels, FP is the number of False Positive voxels, FN is the number of False Negative voxels, ${\textbf{T}_{i}}\in \{0,1\}$ is the true value of the voxel $i$, and $\textbf{P}_{i}\in [0,1]$ is the predicted probability of the voxel $i$. 

The results of the soft-STAPLE algorithm described above is soft ground truth labels. When the ground truth is represented by a soft mask,  i.e., $\textbf{T}_{i}$  is a soft label in the range [0,1], we can still use the same definition  (\ref{dice_loss}) to obtain a soft version of the Dice score. 

\section{Experimental Results}
\label{sec:experimental_results}

We next evaluated the proposed label fusion method on a publicly available MS lesion dataset.  We trained a lesion segmentation FCNN with ground truth masks constructed by several label fusion methods and compared the performance of these methods using  a cross-validation technique.

\begin{figure}[ht]
  \centering
  \includegraphics[width=\linewidth]{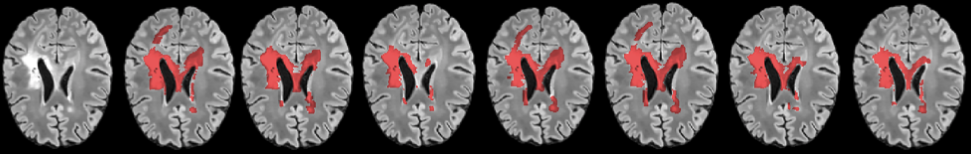}
  \caption{FLAIR modality slice (left image) and its corresponding 7 MS lesions manual delineations. The data is taken from the MICCAI 2016 MS lesions segmentation challenge.}
  \label{fig:manual_delineations}
\end{figure}

{\bf Dataset.} We used the dataset of the MICCAI 2016 MS lesions segmentation challenge (MSSEG dataset) \cite{msseg_commowick}. It consists of 15 cases from 3 different sites and 3 different MRI scanners (Philips Ingenia 3T, Siemens Aera 1.5T and Siemens Verio 3T). Each case consists of 4 series of MRI images, composed of   3D FLAIR, 3D T1-weighted, 3D T1-weighted GADO and 2D PD-/T2-weighted scans. Seven manual delineations were provided for each MS patient case with the experts split over the 3 sites providing MR images (Figure \ref{fig:manual_delineations}).  High inter-rater variability was found  among the experts in terms of the Dice overlap measure.

{\bf Network architecture and training details.}
To demonstrate the effectiveness of the proposed label fusion approach we trained a U-net \cite{unet_ronneberger} based FCNN. Due to the relatively small dataset we reduced the number of network parameters as compared to the original U-Net to prevent over-fitting issue. The input to the network is a concatenation of 5 2D slices, corresponding to the different MRI modalities: FLAIR, T1-weighted, T1-weighted GADO, PD-weighted and T2-weighted. Similar to the U-net, the network architecture we used is divided into two pathways of corresponding layers  which are connected to leverage both high- and low-level features: A contracting path alternates $3 \times 3$ convolution layers and $2 \times 2$ max-pooling layers with stride 2 for downsampling. The expansion path alternates $3 \times 3$ convolution layers and $2 \times 2$ transposed convolution layers. All the convolution layers, except for the last one, are followed by a rectified linear unit (ReLU) \cite{relu_nair}. Activations of the last convolution layer are fed to a sigmoid function that produces a probabilistic segmentation map with values in the range of 0 to 1. The network was trained using the Dice score function (\ref{dice_loss}). 

{\bf Compared label fusion methods.}
First we constructed soft masks from the experts' manual delineations adopting the protocol and optimal set of parameters (the dilation size and soft label were 120\% and 0.3 respectively) as described in \cite{soft_labeling_kats}. We next applied the soft-STAPLE and simplified-soft-STAPLE algorithms on the created soft masks to obtain the  ground truth mask for training. As a baseline we applied the standard STAPLE algorithm on the original binary delineations. Finally, we also constructed a dilated-STAPLE soft masking from the STAPLE masking by applying the protocol that was used to obtain the soft mask for each expert. 

\begin{figure}[ht]
  \centering
  \includegraphics[width=\linewidth]{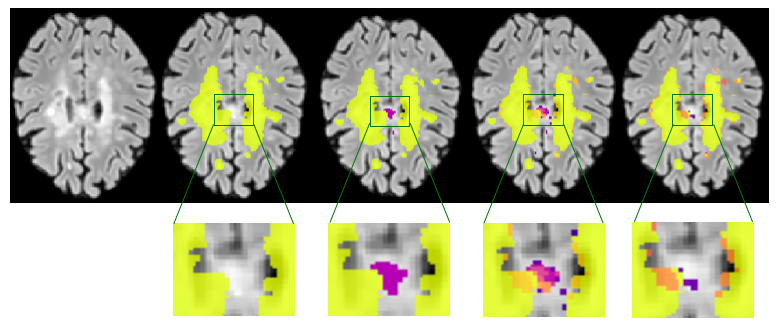}
  \caption{Illustration of consensus masks used as ground truth during FCNN training. Yellow color denotes pixels with  a value of 1, colors gradually changed to violet as the value of the pixels decreases. From left to right: FLAIR modality,  mask crated by STAPLE  and soft masks created by dilated STAPLE, soft-STAPLE and simplified-soft-STAPLE.}
  \label{fig:consensus_masks}
\end{figure}

Examples of the different consensus masks are shown in Figure \ref{fig:consensus_masks}. The proposed soft-STAPLE algorithm benefits from the anatomical knowledge  provided by the conditionally dilated expert' annotations. Consequently the soft consensus mask created by the soft-STAPLE algorithm includes pixels surrounding the lesion from the most likely experts' delineations. We suggest that these pixels contain some additional lesion level information. The simplified-soft-STAPLE algorithm weights the experts' annotations in a different way. We observed that a smaller number of lesion surrounding pixels were included in the soft consensus mask constructed via this method.

{\bf Experiments and results.} We evaluated the proposed methods on the MSSEG dataset by applying a leave-out cross-validation approach. In each fold a set of 3 subjects was used for testing, such that subjects within the set were acquired by a different scanner type. The 5-fold cross-validation results produced the final performance evaluation measures.

Table \ref{tab:experiments} summarizes the performance of the FCNN models that were trained with the compared ground truth masks. 
The test performance shown in Figure \ref{tab:experiments} was evaluated using the constructed STAPLE mask similar to the MSSEG challenge protocol \cite{msseg_commowick}. In addition we evaluated the test results using the ground truth of seven experts: the test image results were separately evaluated with respect to each expert and the average score is reported. 

 The results show that the consensus mask created with the soft-STAPLE algorithm provided valuable information about near-contour voxels during the training phase. The model trained with this mask achieved significant improvement in recall and the highest Dice measure compared to the baseline. The model trained with the mask created by simple-soft-STAPLE also benefited from the additional anatomical information, but achieved a smaller performance gain. This result is consistent with the observation that the mask contains a smaller number of lesion-surrounding pixels.
 The masks that are created using the standard STAPLE algorithm followed by conditionally dilated  contribute less beneficial information to the training process; we believe this is due to the fact   that the dilated region is comprised of  voxels with the same label - same confidence weight to be a lesion - from all experts, regardless of their relative performance.

\begin{table}[ht]
  \caption{Results of training with different consensus ground truth masks.}
  \centering
  \begin{tabular}{|m{3.5cm}|x{1.2cm}|x{1.3cm}|x{1.2cm}|x{1.2cm}|x{1.3cm}|x{1.2cm}|} \hline
  \multirow{2}{*}{Ground Truth}  & \multicolumn{3}{c|}{Experts} & \multicolumn{3}{c|}{STAPLE} \\ \cline{2-7}
  & dice & precision & recall & dice & precision & recall \\ \hline
  STAPLE & 54.3 & 53.3 & 60.4 & 58.3 & 67.1 & 53.0 \\ \hline
  Dilated STAPLE  & 52.2 & 51.9 & 59.3 & 56.0 & 65.1 & 51.6 \\ \hline
  Soft STAPLE & {\bf 56.1} & 53.8 & {\bf 64.0} & {\bf 60.1} & 67.5 & {\bf 55.7} \\ \hline
  Soft STAPLE Simplified & 55.0 & {\bf 54.3} & 60.9 & 59.0 & {\bf 68.4} & 53.4 \\ \hline
  \end{tabular}
  \label{tab:experiments}
\end{table}

To conclude, in this study we proposed a soft-STAPLE  algorithm to generate ground truth labels from a set of manual labels. The  proposed algorithm was tested on the MS lesion segmentation task,with  manual annotations from several experts. We first extended each expert label mask by adding soft labeled voxels which were similar to the annotated voxels in both location and intensity.
Then we applied  the  soft-STAPLE algorithm to obtain an integrated ground truth. We showed that training the FCNN with the computed labels leads to better model generalization and performance gain. The soft-STAPLE concept is general and can be harnessed to improve other medical image segmentation tasks. In this paper the soft labels were obtained by extending the manual labels up to an anatomical border. The soft-STAPLE can be also applied  when the expert is an automatic  probabilistic classifier such as a logistic regression or a neural network.

\comment{
\section*{Appendix: EM with soft observations} Consider a latent variable model $p(x,y;\theta)$ where $x$ is latent, $y$ is observed and $\theta$ is the model parameter. Unlike the standard setting of the EM algorithm, instead of observing $y$ we are only given a distribution $q(y)$.  We want to optimize the soft version of the log-likelihood function: $L(\theta)= \sum_y q(y) \log \sum_x p( x,y)$.It can be easily verified that for every conditional distribution $r(x|y)$:
\begin{equation}
L(\theta) = \sum_y q(y) \log p(y;\theta) \ge \sum_{y,x} q(y)r(x|y) \log \frac{p(x,y;\theta)}{r(x|y)}
\label{lem2}
\end{equation}
and there is  equality when $  r(x|y) = p(x|y;\theta)$. Therefore:
\begin{equation}
\theta_{ML} =  \arg \max_{\theta} \max_{r} \sum_{y,x} q(y)r(x|y) \log \frac{p(x,y;\theta)}{r(x|y)}.
\label{lem3}
\end{equation}
Approximating the double minimization in Eq. (\ref{lem3}) with an alternating minimization  we get the  soft EM algorithm.
In the E-step we compute $\hat{r}(x|y) = p(x|y;\theta)$ and in the M-step we update $\theta$:  
$$
\hat{\theta}=  \arg \max_{\theta} \sum_y q(y) \sum _x  \hat{r}(x|y) \log p(x,y;\theta).
$$
The alternating maximization view of the soft EM algorithm implies that   it  monotonically increases $L(\theta)$.
Once we have found the ML estimation of the model parameters $\theta$ we can reconstruct the hidden random variable $x$:
$$
\hat{p}(x) = \sum_y q(y) p(x|y;\theta).
$$
}


\end{document}